\documentclass[%
 reprint,
 amsmath,amssymb,
 aps,
 jpp,
]{revtex4-2}

\usepackage{graphicx}
\usepackage{siunitx}
\usepackage{dcolumn}
\usepackage[symbol]{footmisc}
\usepackage{hyperref}
\usepackage{epstopdf, epsfig}

\usepackage{bm}

\begin{document}

\preprint{APS/123-QED}

\title{A feasibility study of using X-ray Thomson Scattering to diagnose the in-flight plasma conditions of DT cryogenic implosions}

\author{H. Poole$^{1}$}
 \altaffiliation[Corresponding author address: ]{hannah.poole@physics.ox.ac.uk}
\author{D. Cao$^2$}
\author{R. Epstein$^2$}
\author{I. Golovkin$^3$}
\author{T. Walton$^3$}
\author{S. X. Hu$^{2,4}$}
\author{M. Kasim$^1$}
\author{S. M. Vinko$^{1,5}$}
\author{J. R. Rygg$^{2,4}$}
\author{V. N. Goncharov$^{2,4}$}
\author{G. Gregori$^1$}
\author{S. P. Regan$^{2,4}$}

\address{$^1$Department of Physics, University of Oxford, UK.}
\address{$^2$Laboratory for Laser Energetics, University of Rochester, USA}
\address{$^3$Prism Computational Sciences, USA}
\address{$^4$Department of Mechanical Engineering, University of Rochester, USA}
\address{$^5$Central Laser Facility, STFC Rutherford Appleton Laboratory}

\date{\today}

\begin{abstract}
The design of inertial confinement fusion (ICF) ignition targets requires radiation-hydrodynamics simulations with accurate models of the fundamental material properties (i.e., equation of state, opacity, and conductivity). Validation of these models are required via experimentation. A feasibility study of using spatially-integrated, spectrally-resolved, X-ray Thomson scattering (XRTS) measurements to diagnose the temperature, density, and ionization of the compressed DT shell and hot spot of a laser direct-drive implosion at two-thirds convergence was conducted. Synthetic scattering spectra were generated using 1-D implosion simulations from the LILAC code that were post processed with the X-ray Scattering (XRS) model which is incorporated within SPECT3D. Analysis of two extreme adiabat capsule conditions showed that the plasma conditions for both compressed DT shells could be resolved.
\end{abstract}

\maketitle

\section{\label{sec:intro}Introduction}

The design of inertial confinement fusion (ICF) targets is a challenging task that requires, among others, hydrodynamic simulations with knowledge of the shocked materials' equation of state (EOS) if ignition conditions are to be achieved \cite{Lindl04, Hurricane14, Regan18, Goncharov17, Campbell17}. 
The theoretical modelling of the extreme matter properties reached during the capsule implosion is difficult due to the need of a quantum mechanical treatment of the degenerate electrons, moderate strongly-coupled ions and many-particle correlations \cite{Gaffney18, Hu10, Hu18, Hu17}. 
Uncertainty in the EOS of matter under this regime results in unconfirmed calculations for transport properties, ionization balance, and energy and temperature equilibration \cite{Wang13, Vinko13, Chapman15, White14,Grabowski2020}.
Therefore, experimental validation is vital for benchmarking and developing reduced models that can be implemented in radiation hydrodynamic codes.

At present, the diagnosis of the physical properties of dense plasmas produced in ICF implosions is limited due to the difficulty in achieving the required accuracy and spatial resolutions \cite{Hurricane19, Glebov10, Regan12, Glenzer09} for different model predictions to be tested. Over the past couple of decades there has been a push to develop new diagnostics that may be able to resolve different regions of the imploding capsule, particularly the hot spot, the compressed shell and the coronal plasma. 
Multi-keV spectrally resolved x-ray Thomson scattering (XRTS) is one of these techniques \cite{Glenzer09, Gregori08, Chapman14}.  

The first experimental observation of noncollective, inelastic x-ray scattering from shocked liquid deuterium is discussed in Ref. \cite{Regan12}. This demonstrated the capabilities of inferring the electron temperature, ionisation and electron density from the shapes and intensities of the elastic (Rayleigh) and inelastic (Compton) components in the scattering spectra in ICF dense matter. 
However, the scattering data had no spatial information, nor did the analysis performed provide the capability to separate the contribution from different regions.

Spatial temperature and ionization profiles were determined from a near-solid density foam using a collimated X-ray beam in Ref. \cite{Gamboa14}. This data, produced using the Imaging X-ray Thomson Spectrometer (IXTS) at the Omega laser facility \cite{Boehly1997, Gamboa12}, determined the temperature and ionization state of the carbon foam at multiple positions along the axis of the flow. Good agreement was found between the experiment and theoretical predictions with the exception of the high-temperature, low-density rarefaction region of the blast wave.

Simultaneous collective and non-collective scattering data for dynamically compressed deuterium was collected in Ref. \cite{Davis16} using the $2\,\si{keV}$ Si Ly-$\mathrm{\alpha}$ line. This focused on compression states of $\rho/\rho_0 \sim 2.8-4.05$. The mass density was determined using the VISAR shock velocity using current EOS data. This allowed for a restriction on the parameter space when determining the ionization from the XRTS data.

However, to date, there has been no attempt to field an XRTS diagnostic on a full laser direct-drive ICF implosion. In this report the feasibility of utilising spatially integrated XRTS measurements to determine the in-flight conditions of the compressed DT shell will be investigated. 
The study involved analysing the X-ray scattering data produced by targets with very different adiabats.
The adiabat is defined as the ratio of the plasma pressure to the Fermi-degenerate pressure \cite{SpringerB} and for DT fuel is given by \cite{Craxton15}
\begin{equation}
    \label{adiabat}
    \Gamma_{DT}
    \simeq
    \frac{P_{\mathrm{Shell}}[\mathrm{Mbar}]}{2.2\left[\rho[\mathrm{g/cm^{3}}]\right]^{5/3}}
    \,.
\end{equation}
Confinement properties of an ICF capsule depend on the areal density of the compressed shell and hot-spot, $\rho R$. The areal density is controlled by varying the entropy of the fuel, which is determined by the fuel adiabat. 
For ignition to occur, a large enough areal density (low adiabat), $>\numrange[range-phrase=-]{0.2}{0.5}\,\si{g/cm^2}$, and hot enough core, $\sim5-12\,\si{keV}$, are required \cite{Betti16, AtzeniB}. However, targets imploded on a low adiabat are susceptible to hydrodynamic instabilities \cite{Landen20, Edwards13} that drive the rapid growth of nonuniformities. Therefore, an important part of ICF research involves optimisation of the adiabat \cite{Anderson04, Melvin15, Dittrich14}.
In experiments, however, direct measurements of the in-flight fuel adiabat and densities are not yet achievable, instead they are inferred from the neutron yield and x-ray self-emission \cite{Cerjan13}.

This paper presents dual-channel XRTS as a possible diagnostic to retrieve spatial information on the in-flight conditions of an ICF implosion.
The analysis is performed by constructing synthetic, spatially integrated, spectra using the collision-radiative code SPECT3D \cite{MacFarlane07}, including the x-ray scattering simulator \cite{Golovkin13}, which is a post-processor of the 1-D radiation hydrodynamic code LILAC \cite{Delettrez87}. 

\section{\label{sec:setup}Proposed Experimental Setup}

\begin{figure}[t]
	\centering
	\includegraphics[width=0.45\textwidth]{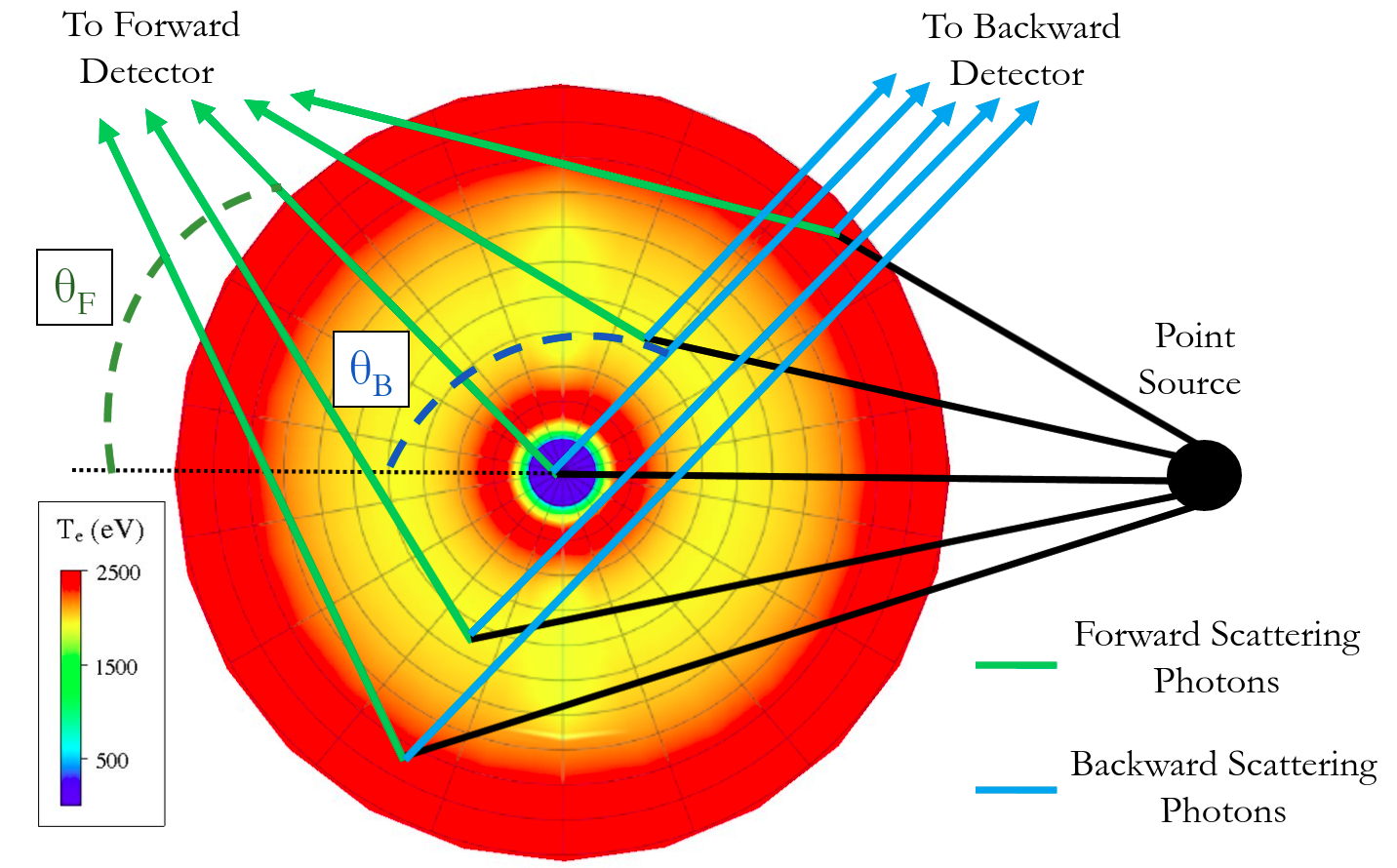}
	\caption{The 3D inferred temperature profile from Spect3D using the 1D simulation data produced by the LILAC code. Schematic of the scattering events, recorded on the detector by SPECT3D, from different zones throughout the implosion are shown. The scattering geometry is demonstrative and not drawn to scale.}
	\label{3d_implosion}
\end{figure}

XRTS is a powerful diagnostic tool for determining the conditions in plasmas where the critical density, $n_c=\epsilon_0m_e\omega^2/e^2$ (where $m_e$ is the mass of an electron, $e$ is the electron charge, $\epsilon_0$ is the electric constant and $\omega$ is the frequency of the laser drive),
exceeds what can be probed by any optical source. The first consideration for an experimental setup, is the power required for the X-ray probe in order to produce a scattering signal that can be observed above background noise. The total number of photons scattered into a detector, $N_d$, can be estimated as \cite{Glenzer09}
\begin{eqnarray}
 	\label{detected_photons}
    N_{d}
    = &&\,
    \left(\frac{E_L}{h\nu}\eta_x\right)\left(\frac{\Omega_{\mathrm{plasma}}}{4\pi}\eta_{\mathrm{att}}\right)\left[\frac{n_e\sigma_{\mathrm{Th}}\ell}{\left(1 + \alpha\right)^2}\right]\nonumber\\
    &&\times
    \left(\frac{\Omega_{\mathrm{det}}}{4\pi}\eta_d\right)\,,
\end{eqnarray}
where $E_L$ is the probe laser energy, $\eta_x$ is the conversion efficiency from the laser energy into the probe X-rays, $\eta_{\mathrm{att}}$ is the attenuation of the probe X-rays through the dense plasma, $\Omega_{\mathrm{plasma}}$ and $\Omega_{\mathrm{det}}$ are the solid angles subtended by the plasma and the detector, respectively, $n_e$ is the electron density, $\alpha$ is the scattering parameter, $\sigma_{\mathrm{Th}}$ is the Thomson scattering cross-section, $\ell$ is the path length of the photons through the plasma, and $\eta_d$ is the detector efficiency.

For the plasma conditions investigated here, the scattering fraction, $n_e\sigma_{\mathrm{Th}}\ell$, is approximately equal to $10^{-4}$, where we have taken representative values for the compressed shell to be $n_e\sim 10^{23}\,\,\si{cm^{-3}}$ and $\ell=50\,\,\si{\mu m}$. This small scattering fraction makes fielding XRTS challenging since the signal can easily be
swamped by significant self-emission from the plasma. 
In this feasibility study we show that a probe laser energy of $1\,\si{kJ}$ is required, as will be discussed in detail below. 

A key benefit of fielding XRTS as a plasma diagnostic, is that XRTS can be split into two scattering regimes, the collective and the noncollective, as determined by the scattering parameter,
\begin{equation}
	\label{scattering_parameter}
	\alpha 
	= 
	\frac{1}{k\lambda_S}\,,
\end{equation}
where $k$ is the scattering vector, and $\lambda_S$ is the screening length. 
In the noncollective regime, the incoming wave `probes' through the screening sphere and the scattering spectrum therefore reflects the electron velocity distribution.
In contrast, the collective scattering regime reflects the collective motion of the electrons.
Designing an experiment where both regimes can be recorded can reduce the error on the inferred plasma parameters. 

\begin{figure*}[t]
	\centering
	\includegraphics[width=0.995\textwidth]{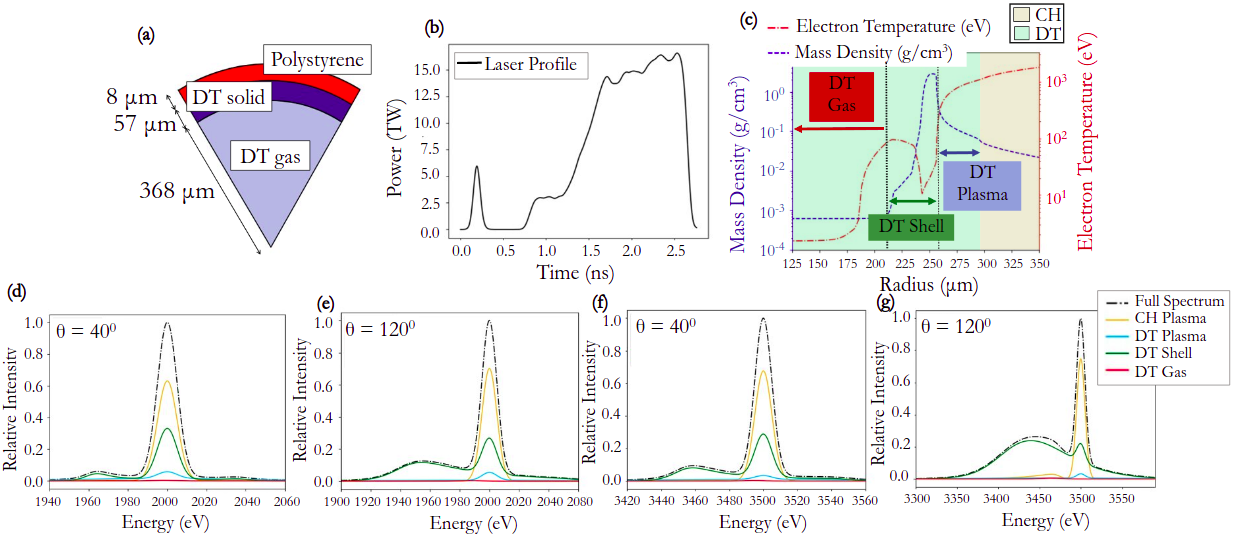}
	\caption{\textbf{(a)} Simulated target design, with an adiabat of $2.8$, fired with laser profile shown in \textbf{(b)}.
	\textbf{(c)} Density and electron temperature conditions in the ICF implosion across the shock wave at two-thirds compression, $t=2215\,\si{ps}$, as determined by the LILAC code for the target. 
	The scattering contributions from the DT in the unshocked fuel, compressed shell and coronal plasma has been isolated and compared to the fully integrated spectrum. 
	For a $2\,\si{keV}$ probe, the contribution from each region of the plasma to the overall scattering spectrum is shown for both the forward ($40\si{^{\circ}}$), \textbf{(d)}, and backward ($120\si{^{\circ}}$) scattering regime, \textbf{(e)}. The same breakdown of the plasma has been performed with a $3.5\,\si{keV}$ energy probe in \textbf{(f)} and \textbf{(g)}.}
	\label{2.8_conditions}
\end{figure*}

\begin{figure*}[t]
	\centering
	\includegraphics[width=0.995\textwidth]{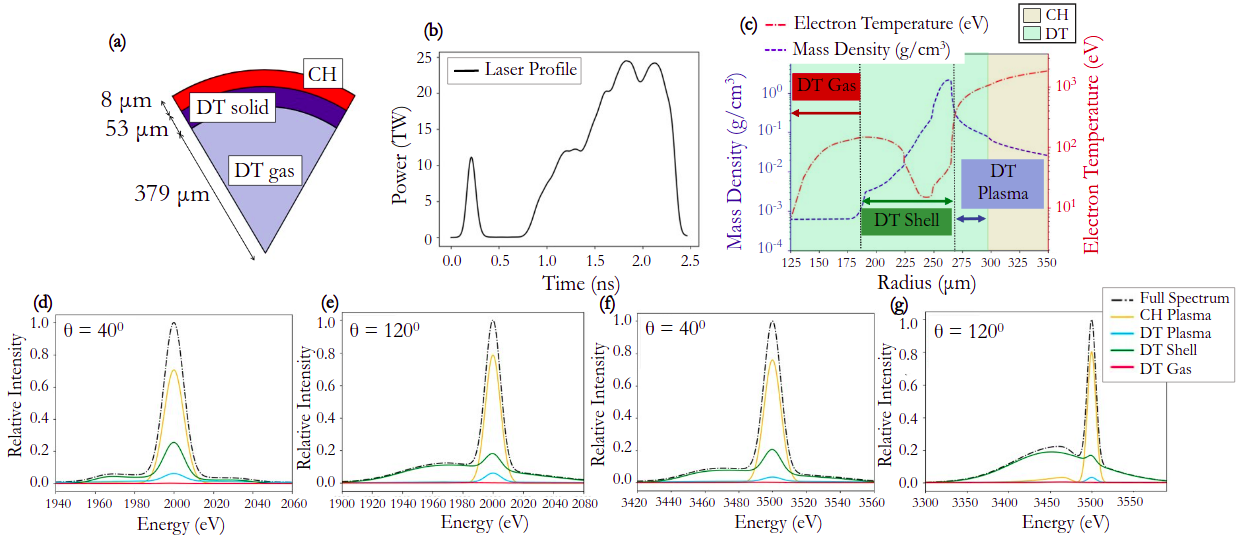}
	\caption{As with figure \ref{2.8_conditions} but with an ICF capsule with an adiabat of $8.0$ and at $t=1901\,\si{ps}$.
	}
	\label{8.0_conditions}
\end{figure*}

To model the X-ray emissivity, a $1\,\si{kJ}$ laser with a $1\,\si{ps}$ pulse length and a source diameter of $100\,\si{\mu m}$ was used to produce a Gaussian X-ray source, with a FWHM of $10\,\si{eV}$, $4.5\,\si{cm}$ away from the imploding target, taking a conservative estimate of $\eta_x=0.01\%$.
Two $12.25\,\si{cm^2}$ charge-coupled device (CCD) detectors were used to collect spectrally resolved radiation. The scattering geometry is shown in Figure \ref{3d_implosion}. The detectors were placed at a distance of $2.4\,\si{m}$ away from the plasma and at scattering angles of $\theta_F = 40\si{^{\circ}}$ and $\theta_B = 120\si{^{\circ}}$. 
The two targets chosen for this investigation are shown in Figures \ref{2.8_conditions} and \ref{8.0_conditions} with adiabats of $2.8$ and $8.0$ respectively.

\begin{figure}[t]
	\centering
	\includegraphics[width=0.455\textwidth]{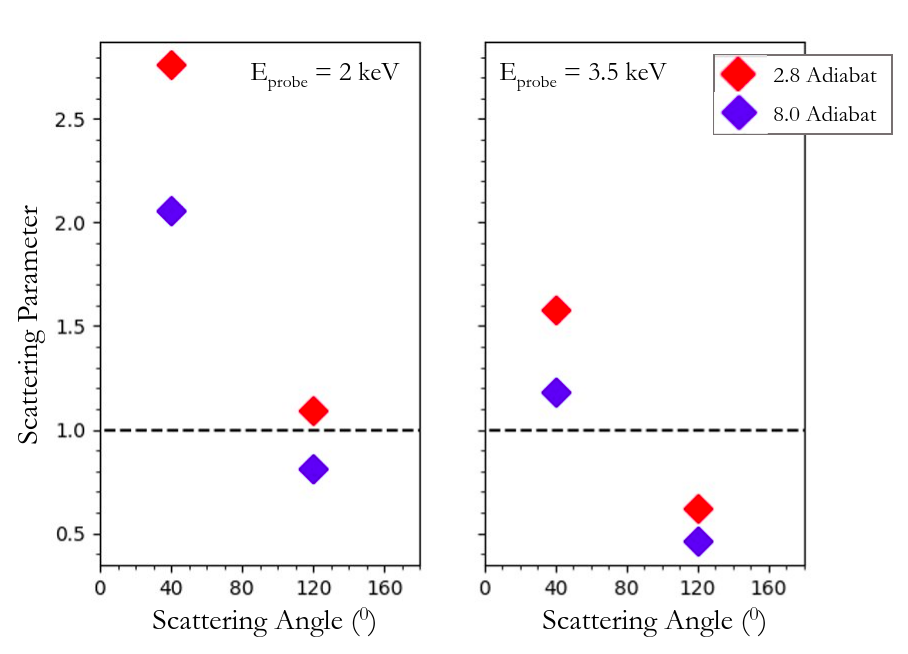}
	\caption{Scattering parameters, $\alpha$, as calculated for the densest zone in the compressed DT shell for each adiabat, scattering angle and probe energy. A dashed line is shown at $\alpha=1$ which is the approximate separation of collective, $\alpha>1$, and noncollective, $\alpha\leq1$, scattering.}
	\label{fig_params}
\end{figure}

Two experimental setups are considered for this paper, one with an X-ray probe energy of $2\,\si{keV}$ and the other using a $3.5\,\si{keV}$ probe. 
The scattering regime recorded by each detector in each setup is shown in Figure \ref{fig_params}. It should be noted that the values for the $\alpha$ parameter shown in the figure are calculated for the densest region in the compressed DT shell, and therefore not representative of the scattering from the ICF capsule as a whole. To determine the scattering signals from each region of the implosion, the fully integrated scattering spectra must be determined.

The plasmon frequency shift for the high adiabat target is $\sim27\,\si{eV}$, which increases to $\sim30\,\si{eV}$ for the low adiabat target. In order to distinguish this plasmon scattering, a narrow band X-ray probe must be used. To achieve this in an experimental setup, the source must be chosen carefully.

Previous experiments have successfully used a crystal imaging system with a Si He$_\alpha$ line at $\sim1.865\,\si{keV}$ \cite{Stoeckl14} to radiograph OMEGA cryogenic implosions \cite{Stoeckl17} but the required x-ray fluence may not be enough. 
Alternatively, Cl K$_\alpha$ at $\sim2.62\,\si{keV}$ or Cl Ly-$\alpha$ at $\sim2.96\,\si{keV}$ could be used, which would require lasers of energy $650\,\si{J}$ and $300\,\si{J}$, respectively \cite{Urry05}.

An important consideration to make before extrapolating this work to an experimental campaign, is predicting the level of noise on the scattering signal. Many factors can contribute to the noise level such as the self-emission, the time-gating of the detector, the detector efficiency \textit{etc.}. For the sake of simplicity in this paper, these parameters will be collected into one function, $G$, which we approximate as $10^{-5}$ \cite{Pak04}. 

\section{\label{sec:spect3d}Obtaining simulated spatially integrated spectra}

\begin{figure}[b]
    \centering
    \includegraphics[width=0.495\textwidth]{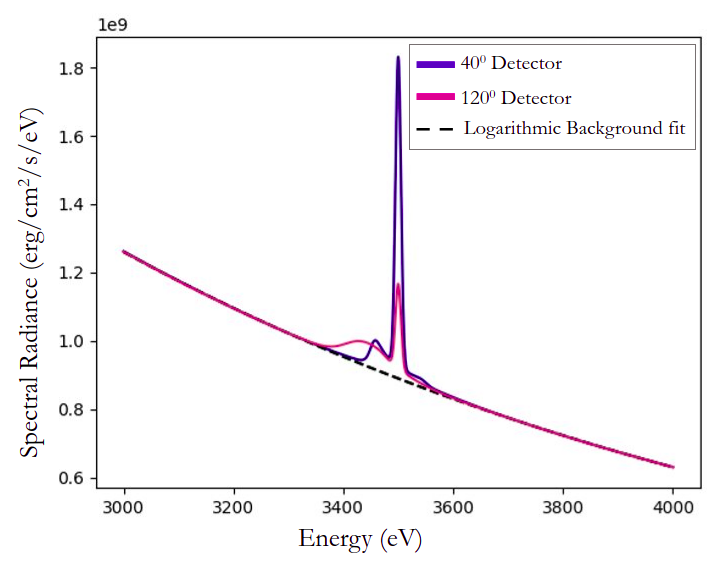}
    \caption{Logarithmic fit to background self-emission for 2.8 adiabat target and a $3.5\,\si{keV}$ X-ray probe. The background self-emission is assumed to fit $A\mathrm{log}^2(I) + B\mathrm{log}(I) + C$, where the constants $A$, $B$ and $C$ are found for each scattering setup.}
    \label{self_emission}
\end{figure}

\begin{figure*}[ht]
	\centering
	\includegraphics[width=0.75\textwidth]{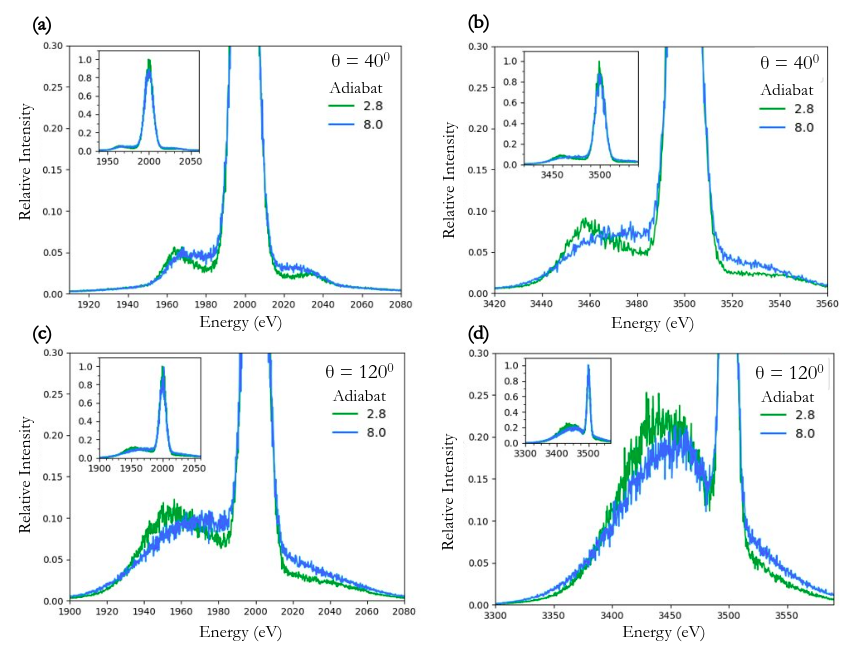}
	\caption{X-ray scattering data produced by Spect3D for LILAC simulations with adiabats of $2.8$ and $8.0$. \textbf{(a)} and \textbf{(b)} Forward scattering spectra for a $2\,\si{keV}$ probe and a $3.5\,\si{keV}$ probe, respectively. \textbf{(c)} and \textbf{(d)} Backward scattering spectra for a $2\,\si{keV}$ probe and a $3.5\,\si{keV}$ probe, respectively.}
	\label{Adiabats}
\end{figure*}

The cryogenic DT implosion plasma conditions were calculated using the LILAC code. The LILAC code is a 1-D spherical lagrangian, radiation-hydrodynamics code \cite{Delettrez87} that simulates symmetric, laser direct-drive implosions.  It includes laser ray-tracing with an inverse bremsstrahlung model that can also account for cross-beam energy transfer \cite{LILAC_CBET}.  LILAC also includes a nonlocal thermal transport model that uses a simplified Boltzmann equation with a Krook collision term \cite{LILAC_nonlocal_TT}, multi-group radiation diffusion, and a first-principles equation-of-state
(FPEOS) model \cite{FPEOS_DT,FPEOS_CD} and opacity (FPOT) model \cite{FPOT_DT} derived from molecular dynamics methods.

In this work, the focus is on the time when the capsule is at two-thirds compression, 
$R_{\mathrm{Ablation\,surface}} / R_{\mathrm{Vapor,\,initial}} = 2/3$. The inhomogenity of the plasma results in different scattering signals from different regions of the plasma. It is paramount that we are able to simulate fully spatially integrated spectra, accounting for opacity and self-emission of the plasma in order to determine, for a given scattering geometry, the dominant scattering features. This provides insightful information to design the experiments.

SPECT3D is a spectroscopy code produced by Prism Computational Sciences which post-processes hydrodynamics code output and simulates high-resolution spectra and images for LTE and non-LTE plasmas in 1-D, 2-D, and 3-D geometries \cite{MacFarlane07}. It computes a variety of diagnostic signatures that can be compared with experimental measurements including: time-resolved and time-integrated spectra, space-resolved spectra and streaked spectra, filtered and monochromatic images, X-ray diode signals. In a SPECT3D simulation, the radiation incident at a detector is computed by solving the radiative transfer equation along a series of lines-of-sight (LOSs) through the plasma grid. At each plasma volume element along a LOS, the frequency-dependent absorption and emissivity of the plasma is calculated. The scattering cross-section is computed using local values of the plasma conditions based on the formalism originally developed in Refs. \cite{Gregori03, Crowley13}. Scattered X-ray photons are added to the local source function, allowing SPECT3D to utilize the same algorithms as it uses for plasma self-emission. It is assumed that the radiation from a non-monochromatic, isotropically emitting point-like X-ray source is scattered within each volume element of the SPECT3D spatial grid. The source is specified by its photon-energy-dependent intensity and location in 3-D space. The intensity of the radiation from the source is adjusted for each volume element based on the distance to the source. It includes attenuation due to plasma absorption and the change in the solid angle. The radiation flux at each pixel in the detector plane is calculated by integrating the scattered radiation along each LOS. The scattering angle is computed for each volume element based on the LOS and the line that connects the volume element center and the source \cite{Golovkin13}.

For this paper, an additional feature was added to the original implementation which allows for certain plasma cells to be excluded from contributing to the scattered signal. This allows for studying the contribution of particular plasma regions to the total scattered spectrum. Models for computing self-emission and absorption coefficients remain the same in each zone regardless of whether the flag for excluding scattered signal is set or not.

The addition of this feature allows spectra from isolated regions of the plasma to be compared to the fully integrated spectra in figures  \ref{2.8_conditions} and \ref{8.0_conditions}. The inelastic scattering in each detector is dominated by the scattering from the compressed DT shell. This gives us confidence that an experiment designed to retrieve scattering spectra at this time during the implosion will be representative of the conditions in the compressed shell.

Using the output from SPECT3D, simulated experimental data was produced by first removing the background noise due to the self-emission of the plasma. A logarithmic fit, was assumed for the background, as shown in Figure \ref{self_emission}. Then random Gaussian noise, with a standard deviation of $7.5\%$ was added to the signal. The resultant spectrum is shown in figure \ref{Adiabats}. Note that we have assumed the noise to be independent of the number of photons per pixel arriving at the detector. In cases where the signal is weak, the signal will be limited by the pixel's photon counts, and in such conditions we would expect the noise to be larger in the wings of the spectrum.
Future work will focus more closely on the noise error analysis for photon-limited signals.

Utilising XRTS to determine the adiabat of an ICF capsule would be a valuable diagnostic development. Figure \ref{Adiabats} demonstrates that for experimental conditions with identical scattering setups, the two extreme adiabat conditions considered here produce notably differing scattering spectra. 
In both the $2\,\si{keV}$ and $3.5\,\si{keV}$ case, the plasmon scattering seen in the forward scattering detector, can be used to determine the difference in electron density between the two adiabats. 
The difference between the inelastic scattering features from the two adiabats seen in figure \ref{Adiabats}\textbf{(c)} is a result of only the low adiabat remaining in the collective scattering regime. The high adiabat's inelastic scattering feature has become dominated by Compton scattering. This is evidenced by the broadening of the inelastic peak and the lose of a forward plasmon shift peak. This change in scattering features is evidence of its higher electron temperature.

\section{\label{sec:results}Results}

\begin{figure*}[t]
	\centering
	\includegraphics[width=0.995\textwidth]{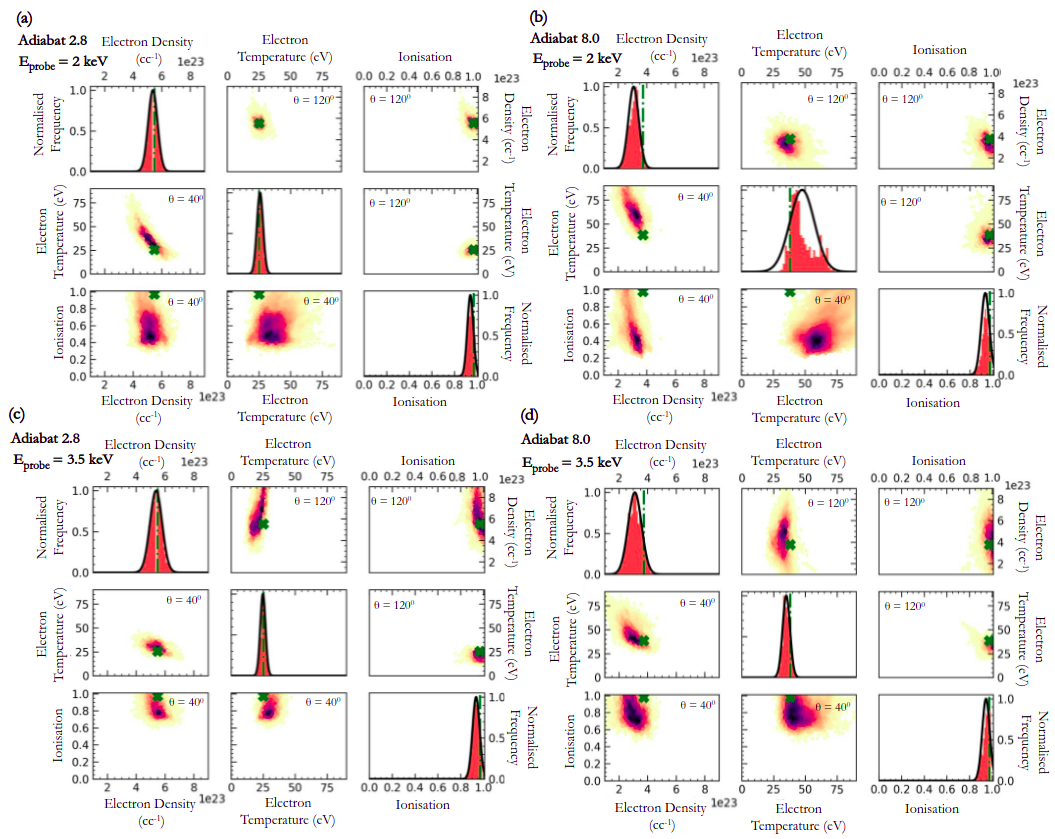}
	\caption{MCMC parameter convergence fitting the entire spectrum using equation \ref{cost_ion} as a cost function. Variation in DT plasma parameters from; \textbf{(a)} $2.8$ adiabat and $2\,\si{keV}$ probe, \textbf{(b)} $8.0$ adiabat and $2\,\si{keV}$ probe, \textbf{(c)} $2.8$ adiabat and $3.5\,\si{keV}$ probe, and \textbf{(d)} $8.0$ adiabat and $3.5\,\si{keV}$ probe. The scatter plots show the correlation between each DT parameter. The lower quadrant scatter plots are taken from the $40\si{^{\circ}}$ scattering data, whilst the upper quadrant shows the $120\si{^{\circ}}$ scattering data. The scatter plots have been coloured to represent the spatial density of points. The diagonal plots show the combined histograms for each parameter from both the scattering regimes. Superimposed on each histogram is a normal distribution of the fits. The mass-averaged parameter values from the LILAC 1-D simulation are highlighted as a green dashed line or cross.}
	\label{ION_MCMC_matrix}
\end{figure*}

\begin{figure*}[t]
	\centering
	\includegraphics[width=0.995\textwidth]{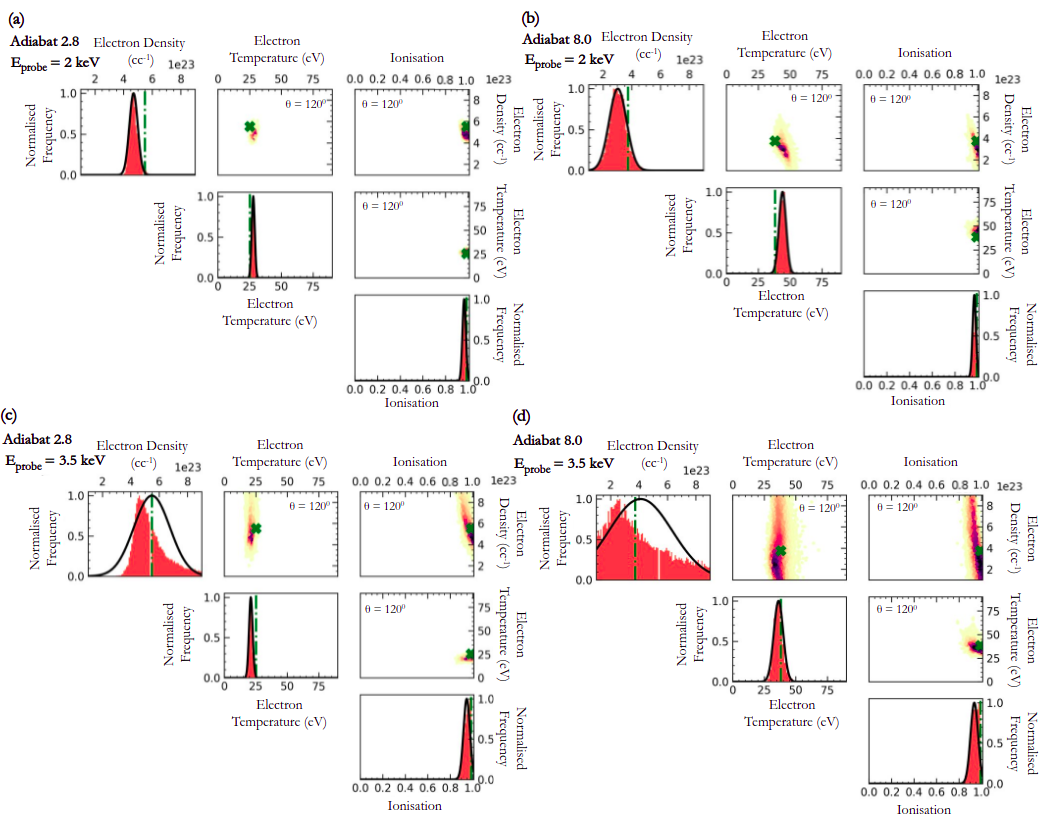}
	\caption{As with Figure \ref{ION_MCMC_matrix} but for MCMC analysis focused solely on the inelastic scattering using equation \ref{cost_single} as a cost function. As discussed in Section \ref{sec:discussion}, the forward scattering analysis has been omitted as the widths of the inelastic scattering features in the forward  scattering  regime  are  small relative  to  the  source FWHM. This means, without the elastic scattering feature, MCMC converges around values not representative of the entire spectra.}
	\label{SINGLE_MCMC_matrix}
\end{figure*}

\begin{table}[h]
    \centering
    \renewcommand{\arraystretch}{1.5}
    \caption{\label{ION_MCMC_results}The full spectral analysis MCMC DT fitting parameters compared to the mass-weighted parameters from the LILAC 1D simulations, focused on the compressed DT shell, for each adiabat and each probe.}
    \begin{ruledtabular}
    \begin{tabular}{l|ccc}
    \textbf{DT Parameter}   & $\mathbf{T_e\,(\si{eV})}$ & $\mathbf{n_e\,(\si{cm^{-3}})}$       & $\mathbf{Z}$      \\ \hline
    \multicolumn{4}{c}{\textbf{Adiabat} $\mathbf{2.8}$}                                                         \\ \hline
    Simulation              & $25$                      & $5.5\times10^{23}$                & $0.97$            \\
    MCMC $2\,\si{keV}$      & $26\pm2$                  & $(5.4\pm0.3)\times10^{23}$        & $0.94\pm0.03$     \\
    MCMC $3.5\,\si{keV}$    & $25\pm2$                  & $(5.4\pm0.4)\times10^{23}$        & $0.94\pm0.03$     \\ \hline
    
    \multicolumn{4}{c}{\textbf{Adiabat} $\mathbf{8.0}$}                                                         \\ \hline
    Simulation              & $38$                      & $3.7\times10^{23}$                & $0.97$            \\
    MCMC $2\,\si{keV}$      & $48\pm9$                  & $(3.1\pm0.4)\times10^{23}$        & $0.94\pm0.04$     \\
    MCMC $3.5\,\si{keV}$    & $35\pm3$                  & $(3.1\pm0.5)\times10^{23}$        & $0.95\pm0.03$     \\
    \end{tabular}
    \end{ruledtabular}
\end{table}

\begin{table}[h]
    \centering
    \renewcommand{\arraystretch}{1.5}
    \caption{\label{SINGLE_MCMC_results} The inelastic scattering analysis MCMC DT fitting parameters compared to the mass-weighted parameters from the LILAC 1D simulations, focused on the compressed DT shell, for each adiabat and each probe.}
    \begin{ruledtabular}
    \begin{tabular}{l|ccc}
    \textbf{DT Parameter}   & $\mathbf{T_e\,(\si{eV})}$ & $\mathbf{n_e\,(\si{cm^{-3}})}$       & $\mathbf{Z}$      \\ \hline
    \multicolumn{4}{c}{\textbf{Adiabat} $\mathbf{2.8}$}                                                         \\ \hline
    Simulation              & $25$                      & $5.5\times10^{23}$                & $0.97$            \\
    MCMC $2\,\si{keV}$      & $28\pm1$                  & $(4.7\pm0.3)\times10^{23}$        & $0.97\pm0.02$     \\
    MCMC $3.5\,\si{keV}$    & $21\pm1$                  & $(5\pm1)\times10^{23}$            & $0.95\pm0.03$     \\ \hline
    
    \multicolumn{4}{c}{\textbf{Adiabat} $\mathbf{8.0}$}                                                         \\ \hline
    Simulation              & $38$                      & $3.7\times10^{23}$                & $0.97$            \\
    MCMC $2\,\si{keV}$      & $44\pm3$                  & $(3.0\pm0.6)\times10^{23}$        & $0.97\pm0.01$     \\
    MCMC $3.5\,\si{keV}$    & $36\pm4$                  & $(4\pm2)\times10^{23}$            & $0.94\pm0.04$     \\
    \end{tabular}
    \end{ruledtabular}
\end{table}

Before extracting the plasma parameters from the spatially integrated simulated spectra, the inverse problem instability must first be addressed, which implies that the same measured spectra could be fitted equally well by very different plasma parameters \cite{Kasim19}. Bayesian inference, using Markov-Chain Monte Carlo (MCMC) to sample the multidimensional space, is a more robust approach to explore the behaviour of the complex multiparameter simulations \cite{Andrieu03}.

Two MCMC explorations that walked through defined parameter spaces to find the ionization, temperature and density that best fit the forward and backward scattering spectra individually are presented in this paper.
The parameter space assumed a uniform distribution with linear sampling for the electron temperature, $1\leq T_e(\mathrm{eV})\leq 1e3$, and ionization, $0\leq Z\leq 1$, whilst taking a logarithmic sampling for the electron density, $1\mathrm{e}20\leq n_e(\mathrm{1/cc})\leq 5\mathrm{e}24$. A large sampling space was used so no bias was placed on the resultant parameters.
One exploration fit the entire spectra, assuming two weighted uniform plasma regions, one containing DT and the other CH.
The cost function used to determine the appropriateness of each MCMC scattering spectra, $I_\mathrm{fit}$, in this case is
\begin{equation}
    \label{cost_ion}
\beta_{\mathrm{cost}} =    \textrm{max}{\left(\frac{I_{\mathrm{fit}} - I_{\mathrm{raw}}}{I_{\mathrm{raw}}}\frac{1}{\sqrt{2}\sigma}\right)^{2}}\,,
\end{equation}
where $I_{\mathrm{raw}}$ is the synthetic scattering spectra shown in Figure \ref{Adiabats} and $\sigma$ is the standard deviation representative of the noise of the synthetic scattering spectra.
In an actual experiment this quantity is not known \textit{a priori} and it must be chosen for MCMC to be able to explore a sufficiently wide parameter space.
For the full spectrum analysis we use $\sigma = 0.075$.
This cost function allowed for equal weighting of the fit to the elastic and inelastic peaks. 

The alternate MCMC analysis focused solely on the inelastic scattering features. As previously discussed, the inelastic scattering for all detectors was dominated by the contribution from the compressed DT shell. Therefore, this MCMC analysis assumed a single uniform plasma condition containing only DT. 
A soft boundary cost function was used in this case as there was no need to apply weighting to the inelastic scattering,
\begin{equation}
   \label{cost_single}
   \beta_{\mathrm{cost}}
   =
    \textrm{max}{\left(\frac{I_{\mathrm{fit}} - I_{\mathrm{raw}}}{\sqrt{2}\sigma}\right)^{2}}\,.
\end{equation}
For this analysis, a value of $0.0005$ was used for $\sigma$ as this is representative of the noise when focused on the inelastic scattering region. 
The forward likelihood of each fit, $P(I_{\mathrm{fit}}|I_{\mathrm{raw}})$, was determined as
\begin{equation}
    \label{logprob}
    P(I_{\mathrm{fit}}|I_{\mathrm{raw}})
    =
    \mathrm{e}^{-\beta_{\mathrm{cost}}}\,.
\end{equation}

To analyse the MCMC data, the DT parameters were plotted on a combined matrix shown in Figures \ref{ION_MCMC_matrix} and \ref{SINGLE_MCMC_matrix}. The scatter plots for each scattering angle are shown separately and have been coloured to represent the spatial density of points.
In Figure \ref{ION_MCMC_matrix}, the histograms along the diagonal are the combined histograms for both the forward and backward scattering parameters. The mean and standard deviation on each parameter was calculated by fitting a normal distribution to the histograms. 

The MCMC parameters were compared to the mass-weighted parameters from the 1-D LILAC simulations. The mass-weighted simulation values were calculated using
\begin{equation}
    \label{mass-weighted}
    \mathrm{\left<F\right>}
    =
    \frac{\sum F_i \rho_i 4\pi r_i^2\mathrm{d}r_i}{\sum\rho_i 4\pi r_i^2\mathrm{d}r_i}\,,
\end{equation}
where $F_i$ is the desired parameter in zone $i$.
The mass-weighted parameters were determined for each region of the implosion. It can be seen in Tables \ref{ION_MCMC_results} and \ref{SINGLE_MCMC_results}, that the MCMC values were in close agreement with the mass-weighted parameters from the compressed DT shell. As discussed previously, this was an expected result, as the high density in the compressed shell meant it dominated the inelastic scattering features.

\section{\label{sec:discussion}Discussion}

There is good agreement between the mass-averaged simulation parameter values and the MCMC distributions. 
The forward scattering fits tend to converge around lower densities, higher temperatures and broader ionisations. This results in either broader or slightly skewed distributions on the DT parameters.
This differing convergence occurs because the ratio between the source FWHM and the width of the inelastic scattering feature in the forward scattering case is very small, particularly for the $2\,\si{keV}$ probe. It would therefore be possible to obtain information on the compressed DT conditions solely using a backward scattering detector. In order to improve the fit in the forward scattering regime, either a narrower bandwidth or a higher energy source should be used. 

The forward scattering spectra have not been used to contribute to the MCMC distributions in Figure \ref{SINGLE_MCMC_matrix}. As the widths of the inelastic scattering features in the forward scattering regime are small relative to the source FWHM, convergence around values representative of the compressed DT shell under these conditions are highly improbable. This exclusion of the forward scattering results in much narrower parameter distributions, particularly for the electron temperature and ionisation.
One reason for the tighter fits is because the low weighting on the CH plasma in the full scattering fits means MCMC assumes the conditions from the uniform DT plasma region are generating both the inelastic and the elastic scattering features. However, from Figures \ref{2.8_conditions} and \ref{8.0_conditions} we can see the Rayleigh scattering is dominated by the conditions in the CH coronal plasma. Therefore, for the electron temperature and ionisation, better convergence on the compressed shell DT conditions is seen when the fitting to the elastic scattering feature is ignored.

In contrast, the full spectrum MCMC analysis produces more statistically accurate results on the electron density. This is due to the inclusion of the predominately collective scattering forward detector. 
Inelastic scattering in the collective scattering spectrum is very sensitive on the electron plasma frequency (which determines its overall shift),
meaning the collective scattering case is best used to determine the electron density. However, it should also be noted that removal of the elastic scattering feature leads to uncertainty as to the magnitude of the inelastic shift. This operation would therefore be difficult to justify under experimental conditions.

Overall the best agreement between the MCMC analysis on the synthetic scattering data and the simulations is obtained with the full spectrum analysis using a $3.5\,\si{keV}$ probe.
Better agreement may be achieved when focusing on the inelastic scattering if a narrower bandwidth probe beam could be used, meaning the forward scattering does not need to be omitted from the analysis.
In fact, this is what it would feasible with a Free Electron Laser \cite{Fletcher2015}.

\section{\label{sec:conclusions}Conclusions}

In summary, spatially-integrated XRTS spectra for 1-D LILAC simulated conditions of low- and high-adiabat, DT cryogenic implosions have been calculated at two-thirds convergence.  
Markov-Chain Monte Carlo analysis was performed for two different scattering setups. 
Information on the compressed shell conditions was obtained as it has been shown to be possible to use the spectral resolution in a spatially integrated measurement to discriminate between different regions in the plasma.
Fielding two detectors, one in the collective and one in the noncollective scattering regime, produced the best agreement with the compressed shell mass-averaged parameters from the simulation.
This technique can be used to resolve both the low- and high-adiabat implosions. 

In the future, similar analysis will be performed on the conditions at stagnation as well as investigations into 2-D and 3-D simulations using DRACO \cite{Radha05} and ASTER \cite{Igumenshchev16}.

\section{\label{sec:results}Acknowledgements}

This material is based upon work supported by the Department of Energy National Nuclear Security Administration under Award Number DE-NA0003856.

M.F.K. and S.M.V. acknowledge support from the UK EPSRC grant EP/P015794/1 and the Royal Society. S.M.V. is a Royal Society University Research Fellow. 

\section{\label{sec:statement}AIP Publishing Data Sharing Policy}

The data that supports the findings of this study are available from the corresponding author, H.P., upon reasonable request.
\bibliography{biblio}

\end{document}